\documentclass[useAMS,usenatbib]{mn2e}
\usepackage{graphicx}

\def\lta{\lower2pt\hbox{$\buildrel {\scriptstyle <}
\over {\scriptstyle\sim}$}}
\def\gta{\lower2pt\hbox{$\buildrel {\scriptstyle >}
\over {\scriptstyle\sim}$}}
\def\pts{\lower2pt\hbox{$\buildrel {\scriptstyle\propto}
\over {\scriptstyle\sim}$}}
  
\def\L{{\mathcal L}}
 \def\epsilon{\varepsilon}
\def\ie{i.e.}

\def\apj{ApJ}
\def\nat{Nature}
\def\apjl{ApJ}
\def\mnras{MNRAS}

\begin{document}

\title[Reverse Shock Emission as a Probe of GRB Ejecta]
{Reverse Shock Emission as a Probe of GRB Ejecta}
\author[E. McMahon, P. Kumar and T. Piran]{E. McMahon,$^1$ P.
Kumar$^1$ and T. Piran$^{2,3}$  \\
$^1$Department of Astronomy, University of Texas, Austin, TX 78712, USA\\
$^2$Racah Institute for Physics, The Hebrew University, Jerusalem,
91904, Israel\\
$^3$Theoretical Astrophysics, Caltech, Pasadena, CA 91125, USA}

\maketitle
\begin{abstract}
We calculate the reverse shock (RS) synchrotron emission in the
optical and the radio wavelength bands from electron-positron pair
enriched gamma-ray burst ejecta with the goal of determining the pair
content of GRBs using early time observations. We take into account an
extensive number of physical effects that influence radiation from the
reverse-shock heated GRB ejecta.  We find that optical/IR flux depends
very weakly on the number of pairs in the ejecta, and there is no
unique signature of ejecta pair enrichment if observations are
confined to a single wavelength band. It may be possible to
determine if the number of pairs per proton in the ejecta is $\gta
100$ by using observations in optical and radio bands; the ratio of
flux in the optical and radio at the peak of each respective
reverse-shock light curve is dependent on the number of pairs per
proton.  We also find that over a large parameter space, RS emission
is expected to be very weak; GRB 990123 seems to have been an
exceptional burst in that only a very small fraction of the parameter
space produces optical flashes this bright.  Also, it is often the
case that the optical flux from the forward shock is brighter than the
reverse shock flux at deceleration. This could be another possible
reason for the paucity of prompt optical flashes with a rapidly
declining light curve at early times as was seen in 990123 and 021211.
Some of these results are a generalization of similar results reported
in Nakar \& Piran (2004).
\end{abstract}

\section{Introduction}
Since the discovery of the bright optical flash from GRB 990123
\citep{akerlof99}, many authors have considered this early
afterglow emission as being due to the reverse shock passing
through baryonic ejecta from the explosion, as was predicted by
\citet{sari99}. If this early emission is indeed caused by the
reverse shock (hereafter RS) passing through the ejecta, then this
emission provides direct information about the composition and
magnetic field strength and orientation in the material ejected
from the inner engine, giving us clues about the nature of the
outflow and the inner engine itself. Early optical afterglow data
from GRBs 990123 \& 021211 have already been used to infer that
the ejecta from these GRBs may have been fairly highly magnetized
baryonic ejecta \citep{zhang03,fan02,kp03,mkp04}.

Only the observations of GRBs 990123 \& 021211 have exhibited this
steeply falling off early optical emission, while other bursts
observed in the optical so quickly following the burst have exhibited
either rising emission (030418), shallower fall-off (021004), or no
emission at all \citep{kehoe01}.  There have also been some {\it
Swift}-detected bursts recently that have not exhibited the steeply
falling off emission expected from the RS \citep[e.g. GRB 050319;
][]{rykoff05} or optical/IR emission varying on very short timescales,
seemingly following the activity of the central engine \citep[GRB
041219a; ][]{blake05,vestrand05}\footnote{Note however, that the
situation in GRB 041219a is not clear as the correlation reported in
\citet{vestrand05} should have been enhanced with a further temporal
division of the optical signal.}. There have also been some early
optical upper limits from Swift of $V \sim 21$ magnitude within
minutes after the burst \citep[e.g. 050219a; ][]{schady05}. This lack
of observations of the optical flash begs the question: are GRBs
990123 and 021211 unique?  What is the reason for the lack of optical
flashes? These early upper limits on RS emission can provide
information about the ejecta and help to answer these questions.

In our previous paper (\citealt{mkp04}, hereafter MKP04), we estimated
the RS emission and break frequencies at deceleration for a baryonic
outflow for 10 bursts with burst parameters determined from afterglow
modeling (assuming deceleration time is on the order of the burst
duration, as occurs in 990123).  We determined that a possible reason
for weak RS optical emission after deceleration for most bursts is due
to a cooling frequency below the optical band at deceleration.
Although the reverse shock emission is expected to be bright ($R$-mag
$\sim 10$) at deceleration, when the cooling frequency drops below the
observing band, no more emission from the ejecta is observed, with the
exception of off-axis emission which falls of very quickly, as $\sim
1/t^{\left(2 + p/2\right)} \sim 1/t^3$ \citep[][]{kp00}.  However, now
that is is possible to quickly follow up GRBs with {\it Swift} UVOT,
we are not seeing any bright optical flashes with this $1/t^3$
falloff.  This suggests that the RS flux is suppressed for some
reason.

In the calculation in MKP04, we assumed the ejecta was purely
baryonic.  This may not be the case.  If many e$^{\pm}$ pairs are
present in the ejecta (mixed with baryons), either ejected from the
source or created by dissipation during the GRB, or if the magnetic
field strength in the ejecta is very large (perhaps because the
outflow was Poynting flux dominated), the RS emission could be
suppressed significantly from the baryon dominated picture.
Pair-enrichment of the ejecta causes reverse shock emission to be
fainter than that expected from completely baryonic ejecta, and a very
high ejecta magnetic field is likely to weaken the reverse shock as
well; both scenarios provide alternate possibilities for suppression
of reverse shock emission.  It is useful then to determine the
defining characteristics of reverse shock light curves and spectra for
a baryonic, lepton-enriched, or highly magnetized ejecta, in order to
distinguish between these possibilities when the { \it Swift}
satellite and other robotic telescopes accumulate more multi-wavelength
afterglow data at early times.

Pair enrichment of the ejecta has been looked at by \citet{li03} and
the case of highly magnetized ejecta has been investigated by
\citet{zk05}.  In this paper, we take another look at pair-enrichment;
we take into account that the reverse shock is mildly relativistic,
and also carry out a self consistent cooling and self absorption
calculation, which is necessary, since the ejecta self absorption
frequency can be on the order of or greater than the cooling
frequency.  \cite{li03} predicted a bright IR flash with pair-enriched
ejecta, as bright as that expected in the optical with strictly
baryonic ejecta; we find that the optical/IR flux levels at
deceleration have very little dependence on the number of pairs. Also,
any change in spectral peak frequency or optical/IR flux at
deceleration due to pair enrichment can also arise if the microphysics
parameters in the ejecta are changed, thus creating a highly
degenerate problem.

We first describe our model of reverse shock emission and the new
physics we have added to the RS emission calculation (Section 2), then
discuss our results (Section 3).  In Section 4, we summarize the
differences between baryonic and pair-enriched ejecta.

\section{Description of The Reverse Shock Model}
Here, we briefly describe our RS model, then focus on the improvements
we have made to the calculation of emission from the RS.  For this
calculation, we use the equations for the ejecta dynamics and RS
emission described in Kumar \& Panaitescu 2003 (KP03), Panaitescu \&
Kumar 2004 (PK04), and Nakar \& Piran 2004 (NP04) and include
synchrotron and inverse Compton cooling.  We have added several new
components to the calculation, namely (1) the effect of synchrotron \&
self inverse-Compton scattering on electron cooling \& synchrotron
self absorption frequencies, (2) inverse Compton cooling of RS
electrons by synchrotron flux of the forward shock (hereafter FS), (3)
absorption of RS photons (particularly in the radio band) in the FS
region as they traverse through the FS on their way to the observer,
and (4) pair-enriched ejecta.

\subsection{The Standard Model: Dynamics \& Synchrotron Emission in
  the RS }
\label{standard} The ejecta dynamics are determined by assuming an
initial Lorentz factor $\Gamma_0$, the burst duration in the
observer frame $t_{GRB}$, number density of the external medium
$n_0=A R^{-s}$ where $R$ is the radial distance from the center of
the explosion, and isotropic equivalent energy $E_{52} = E/10^{52}
$ergs. The Lorentz factor of the shocked ejecta with respect to the
unshocked external medium is given by (PK04)
\begin{equation}
\Gamma
 = \Gamma_0 \left(1 + 2\Gamma_0 \left(n_0/n_{ej}\right)^{1/2}
 \right)^{-1/2}
\end{equation}
where $n_{ej}=E/\left(4 \pi m_p c^2 R^2 \Gamma_0^2 \Delta\right)$ is
the comoving ejecta density and $\Delta$ is the lab frame ejecta
width, taken to be $ct_{dur}+R/\left(2 \Gamma_0^2\right)$, where
$t_{dur}$ is the host galaxy frame burst duration,
$t_{GRB}/(1+z)$.  When $\Delta \sim ct_{dur}$ at deceleration, we
define this case to be the non spreading ejecta case (also called
thick ejecta), whereas when $\Delta \sim R/\left(2 \Gamma_0^2\right)$,
we define this case to be the spreading ejecta case (also called thin
ejecta).

When $\Gamma_0^2 n_0/n_{ej} \gg 1/4$ near shock crossing time (the
time at which the RS reaches the rear of the ejecta shell), the RS is
relativistic and the bulk Lorentz factor $\Gamma \propto R^{(s-2)/4}$
and the ejecta radius $R \propto t^{2/(4-s)}$ ($t$ is time in the
observer frame), prior to deceleration time, for the case of
non-spreading ejecta.  In the case of spreading ejecta, $\Gamma_0^2
n_0/n_{ej} \sim 0.3$ for $s=0$ and $\sim 0.5$ for $s=2$ at shock
crossing time, so the RS is mildly relativistic.  The radial distance
of the ejecta from the center of the explosion at the time when the RS
reaches the back of the ejecta shell (the shock crossing radius) is
given in Table~1, as are the Lorentz factor and observer time at this
radius.  See Table 2 for the scalings of variables with observer time
before deceleration, numerically determined for the mildly
relativistic case.

The reverse shock speed with respect to the unshocked ejecta as
measured in the lab frame is (KP03)
\begin{equation} \label{betaRS}
\beta_0 - \beta_{RS} =
{1.4
\over \Gamma_0^2} \left(\Gamma_0^2 n_0 \over n_{ej} \right)^{1/2}
\end{equation}
where $\beta_{RS}$ is the reverse shock speed measured in the lab
frame and $\beta_0$ is the outflow speed measured in the lab frame as
well.  The above equation is valid in all cases, from nonrelativistic
to relativistic RS cases.  It can be shown (PK04) that the radius at
which the reverse shock has traversed the ejecta is within a factor of
$\sim 2$ of the deceleration radius for both a wind and homogeneous
external medium.

The rate at which the ejecta electrons are swept up by the RS is
given by
\begin{equation} \label{dNe}
{d N_e \over dR } = {1.4 \over \Gamma_0} \left( {n_0 \over n_{ej} }
  \right)^{1/2} {N_{p} \over \Delta}
\end{equation}
where $N_{p}=E/\left(\Gamma_0 m_p c^2\right)$ is the total number of
protons in the ejecta.  For spreading ejecta, $N_e(R) \propto R^{3/2}$
for $s=0$ and $R^{0.68}$ for $s=2$ (determined numerically) and for
ejecta whose width is dominated by the GRB duration $N_e(R) \propto
R^{(4-s)/2}$.

After deceleration (shock crossing time), the ejecta bulk Lorentz factor
is assumed to evolve adiabatically following the Blandford-McKee
solution, as described in PK04, as
\begin{equation} \label{bulkEjecta}
\Gamma_{ej} \propto R^{-(3-s)/2} \left(R^{(2a-3)}
  \Delta'^a\right)^{-1.5(4-s)/(17-4s)}.
\end{equation}
where $a$ is the adiabatic index, set to $4/3$ for relativistic ejecta
and $5/3$ for non-relativistic ejecta, and $\Delta'=\max(\Gamma_{ej} c
t_{dur}, R/2 \Gamma_{ej})$ is the comoving frame ejecta width.  We
take the ejecta expansion in the radial direction to be relativistic
when the minimum electron thermal Lorentz factor in the ejecta $\gta
10$, and nonrelativistic when it is $< 10$.

Synchrotron radiation from the shocked electrons is calculated
assuming a power law distribution of electron energy with index $p$,
i.e. $d N_e/d \gamma \propto \gamma^{-p}$ for $\gamma > \gamma_{ir}$
(where $\gamma_{ir}$ is the minimum electron thermal Lorentz factor
averaged over all of the electrons as they cross the RS front),
$\epsilon_{Br}$, the fraction of energy in the magnetic field, and
$\epsilon_e'$, the minimum thermal energy given to the electrons;
$\gamma_{ir} = \epsilon_e' \left< e_p \right>/m_e c^2$, where $e_p =
m_p c^2 \left(\Gamma' - 1\right)$ is the minimum thermal energy per
proton just behind the RS front, $\left< e_p\right> \sim 0.5 e_p$, and
\begin{equation} \label{gammap}
\Gamma' = \Gamma_0 \Gamma \left( 1 - \beta_0 \beta \right)
  \simeq 0.5 \left({\Gamma \over \Gamma_0}
  + {\Gamma_0 \over \Gamma}\right)
\end{equation}
is the relative Lorentz factor of the shocked and unshocked
ejecta. Since $\Gamma' \lta 2$, we cannot approximate the evolution of
$\gamma_{ir}$ by assuming $\gamma_{ir} \propto \Gamma'$; we
numerically calculate the evolution for radii near the shock crossing
radius, the results given in Table~2.  The comoving magnetic field
strength in the ejecta is given by $B' = \left[8 \pi \epsilon_{Br}
n_{ej} \left<e_p\right> \left(4\Gamma'+3\right)\right]^{1/2}$.

We note that $\gamma_{ir}$ can be less than 1, since $\gamma_{ir} m_e
c^2$ is defined to be the minimum thermal energy for electrons. We
take into account that some fraction of the electrons are
nonrelativistic and emit cyclotron radiation.  We remove these
electrons from the electron column density used for our synchrotron
calculation, i.e. $n_{col,r} = N_e (\gamma_{ir})^{p-1} / 4 \pi R^2$,
and set the minimum thermal energy for electrons to be $m_e c^2$.

After deceleration, for adiabatically cooling electrons, $\gamma_{ir}
\propto V'^{-\left(a_e - 1\right)} \propto \left(R^2
\Delta'\right)^{-1/3}$, where $V'$ is the comoving ejecta volume, and
the ratio of specific heats is $a_e = 4/3$.  The magnetic field is
assumed to be predominantly transverse and frozen into the ejecta,
decaying as $B' \propto \left(R \Delta'\right)^{-1}$. When radiative
cooling becomes less important than adiabatic cooling, the Lorentz
factor of electrons cooling on a dynamical timescale, $\gamma_{cr}$
(discussed in more detail below), evolves in the same manner as
$\gamma_{ir}$.

The observer frame synchrotron injection frequency is calculated with
\begin{equation} \label{nuir}
\nu_{ir} = {0.98 q B' \gamma_{ir}^2\Gamma\over 2\pi m_e c(1+z)} ,
\end{equation}
where we use here and elsewhere the common notation of
$\nu_x$ being the synchrotron frequency of an electron with a
Lorentz factor $\gamma_x$.  The synchrotron flux at the peak of
the $f_{\nu}$ spectrum is given by
\begin{equation}
 F_{pr} = {N_e P_{\nu_p}\Gamma(1+z)\over 4\pi d_L^2}
\label{fnu}
\end{equation}
where $N_e \equiv N_e(R)$ is the number of electrons heated by the
RS determined from Equation \ref{dNe}, $P_{\nu_p} = {1.04 q^3 B/
m_e c^2}$ is the comoving power radiated per electron per unit
frequency at the peak of $f_{\nu}$, and $d_L=
2c\sqrt{1+z}[(1+z)^{1/2} -1]/H_0$ is the luminosity distance.  We
use $H_0 = 65$ km s$^{-1}$ Mpc$^{-1}$ and for simplicity $\Omega =
1$, $\Lambda=0$.  The factors of 0.98 in Equation \ref{nuir} and
1.04 above for $P_{\nu_p}$ are from \citet{wijers99} for the case
of $p=2$.

\begin{table*}
 \vbox to220mm {\vfil 
\includegraphics[height=\textheight,angle=180]{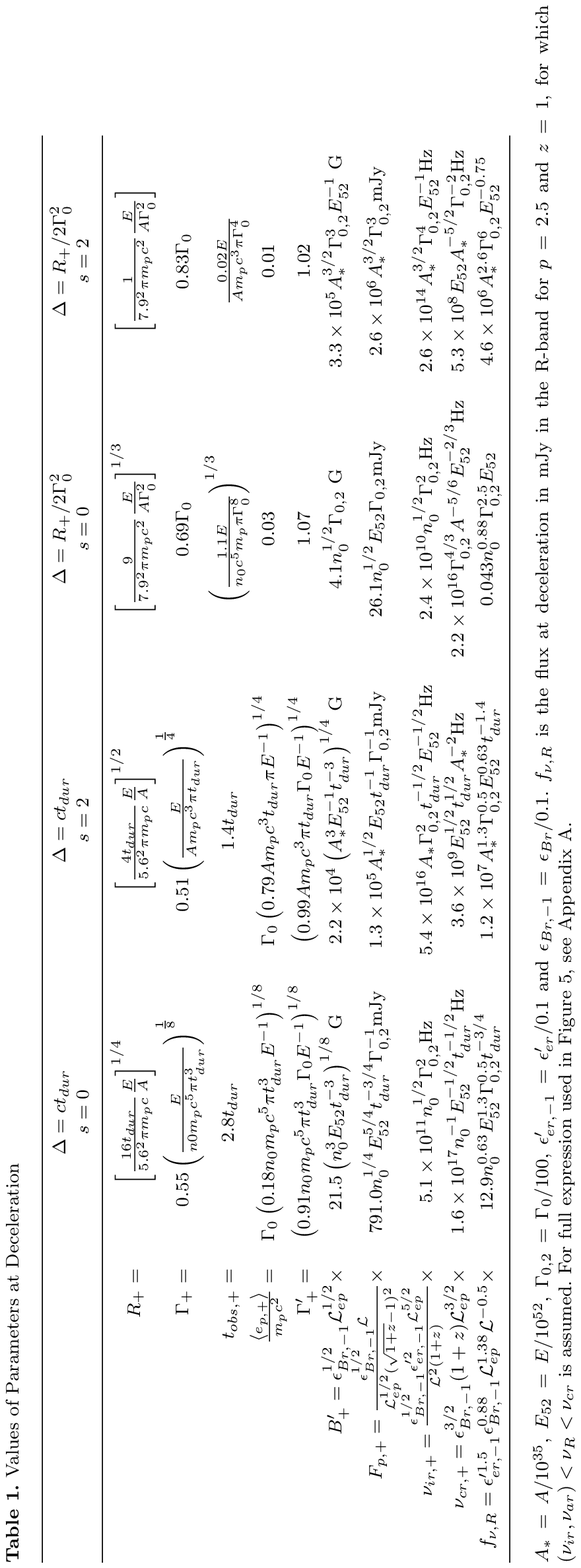}
 \caption{}
\vfil}
 \label{table1}
\end{table*}

\begin{table*}
 \begin{minipage}{0.5\textwidth}%{300mm}
 \caption{Scalings Before Deceleration}
 \begin{tabular}{rcccc}
 \hline & $\Delta=ct_{dur}$ & $\Delta=ct_{dur}$ &
$\Delta=R_+/2\Gamma_0^2$ & $\Delta=R_+/2\Gamma_0^2$ \\
&$ s=0$ & $
s=2$& $ s=0$ & $ s=2$\\
\hline
 $R $ &$t^{1/2}$ & $t$ &$t^{0.67}$&$t^{0.88}$\\
$\Gamma$&$t^{-1/4}$&$t^0$&$t^{-0.17}$&$t^{-0.066}$\\
$\gamma_{ir}$&$t^{0.45}$&$t^0$&$t^{1.7}$&$t^{0.62}$\\
$f_{\nu,X}$ &$t^{-0.18+0.19p} $&$t^{1-0.5p}
$&$t^{1.5\left(p-1\right)}$&$t^{0.012+0.082p}$ \\
$f_{\nu,R}$ & $
t^{0.29+0.19p}$&$t^{-0.5\left(p-1\right)}$&$t^{-0.86+1.5p} $&
$t^{-0.56+0.082p}$\\
 \hline
 \end{tabular}
 \medskip\\ Values of exponents in this table are determined
numerically for mildly relativistic spreading ejecta and determined
analytically for relativistic non spreading ejecta, near, but prior
to, deceleration time.  $t$ is time measured in the observer frame.
For the non-relativistic case of $\Gamma_0^2 n_0/n_{ej} \ll 1/4$,
$\Gamma \propto R^0$ and $R \propto t$.  $ f_{\nu,X}$ is the
synchrotron flux for $\nu_{obs} > (\nu_{ir},\nu_{cr},\nu_{ar})$, and
$f_{\nu,R}$ is the synchrotron flux for
$\left(\nu_{ir},\nu_{ar}\right) < \nu_{obs} < \nu_{cr}$.  Scalings
after deceleration can be found in e.g.  PK04. Note: it is possible
for $f_{\nu,X}$ to be decaying before deceleration because it peaks
well before deceleration.
 \end{minipage}
 \label{table2}
\end{table*}

\subsection{Inverse Compton Cooling \& Synchrotron Self Absorption}
\label{ICabs} We include the inverse Compton process to calculate
electron cooling. The Compton $Y$ parameter is obtained by solving
the equation describing radiative loss of energy for a single
electron with a Lorentz factor $\gamma_e$
\begin{displaymath}
{d \left(\gamma_e m_e c^2\right) \over d t'} = -{\sigma_T c B'^2
  \gamma_e^2 \beta_e^2 \over 6 \pi}  \times
\end{displaymath}
\begin{equation} \label{radloss}
\hfill \left[{f'_{\nu' > \nu'_a} \over f'_{total}} +
  {\tau_{es} \gamma_p^2 \beta_p^2 \over \nu'_p \sigma_T} \int d\nu'
  {f\left(\nu'\right) \sigma_{KN}
  \over \left(1+\tau_{sa}\left(\nu'\right)\right)}   \right]
\end{equation}
where primes denote variables measured in the rest frame of the
shocked fluid.  The bracketed terms on the RHS are effectively
$\left(1+Y\right)$; the first part, the fraction of energy emitted
from one electron with Lorentz factor $\gamma_c$ that is not absorbed
by material in emitting region, is given by
\begin{displaymath}
{f'_{\nu' > \nu'_a} \over f'_{total}} =
\end{displaymath}
\begin{equation} \label{one}
\left\{ \begin{array}{ll} {9x^{4/3}
  - 25.63 \over 5x^{4/3}
  - 25.63 } &
  \textrm{for } \nu'_c > \nu'_a \\
{3\left[2 e^{-x} \sqrt{x} + \sqrt{\pi} \left(1-
  \mathrm{Erf}\left(\sqrt{x}\right)\right) \right] \over
  e^{-x} \left(2 x^{3/2} + 6 \sqrt{x}\right) + 3 \sqrt{\pi}\left(1-
  \mathrm{Erf}\left(\sqrt{x}\right)\right)}  &
  \textrm{for } \nu'_c < \nu'_a
  \end{array} \right. .
\end{equation}
where $x=\nu'_a/\nu'_c$ and Erf is the error function.  This equation
is necessary for calculating electron cooling, since $\nu_c'$ can be
less than the self absorption frequency, $\nu_a'$, in the RS.

The second part of the bracketed term in Equation~\ref{radloss} is
equivalent to $Y$; $\tau_{es}$ is the Thomson optical depth to
electron scattering, $\gamma_p=\min(\gamma_i,\gamma_c)$ and $\beta_p$
is the corresponding velocity in units of $c$,
$\nu'_p=\min(\nu'_i,\nu'_c)$, $\sigma_{KN}$ is an approximation of the
Klein-Nishina correction to the electron scattering cross section,
$\tau_{sa}\left(\nu'\right)$ is the optical depth to synchrotron self
absorption, and $f\left(\nu'\right)$ is the normalized set of broken
power-laws describing the synchrotron spectrum (\ie
$f\left(\nu_p'\right) =1$).

Both of the bracketed terms of the effective $(1+Y)$ (Equation
\ref{radloss}) are dependent on $\nu'_c$ and $\nu'_a$, and in turn,
$\nu'_c$ is also dependent on $(1+Y)$, so the equations for the three
variables must be solved simultaneously; this has been done for all
numerical results in this paper.  It is, however, possible to
semi-analytically estimate $\nu'_c$ when we assume that $Y \ll 1$
and IC does not contribute to the cooling calculation (see Table~1).

The comoving frame synchrotron self absorption frequency is calculated
by equating the comoving synchrotron flux at $\nu'_a$ to flux from
a black body in the Rayleigh-Jeans part of the spectrum, or
\begin{equation} \label{nua}
2 k T {{\nu'_{a}}^2 \over c^2} = f'_{\nu}\left(\nu'_{a}\right)
\end{equation}
where $k T =
\mathrm{max}\left[\gamma_{a},\min(\gamma_{i},\gamma_{c})\right] m_e
c^2$.  A frequent arrangement of the RS break frequencies at
deceleration is $\nu'_{i} < \nu'_{a} < \nu'_{c}$; in this case,
\begin{equation} \label{nua2}
\nu'_{a} = \left[{f'_p \over 2 m_e} {\nu'_{i}}^{(p-1) \over 2}
  \left({q B' \over 2 \pi m_e c}\right)^{1/2} \right]^{2\over(4+p)}
\end{equation}
where $f_p' = N_e \sqrt{3} q^3 B'/\left(4 \pi R^2 m_e c^2\right)$ is the
comoving flux at the peak of $f'_{\nu}$.

The scalings of RS optical and X-ray flux with observer time just
before deceleration for the four cases of $s=0,2$ and $\Delta=(c
t_{dur},R/2 \Gamma_0^2)$ are shown in Table 2.  Only synchrotron
emission is included for these scalings (inverse Compton is also
important in the X-ray), and $Y \ll 1$ is assumed (the scaling for
$f_{\nu,R}$ is not affected, but $f_{\nu,X}$ may be).

After deceleration, if ejecta is in the radiative regime, we continue
to calculate the RS cooling and synchrotron self absorption
frequencies $\nu_{cr}$ and $\nu_{ar}$ and the effective $(1+Y)$ by
solving Equations~\ref{radloss} \& \ref{nua} simultaneously, as done
before deceleration.  If adiabatic cooling is more efficient,
$\gamma_{cr}$ decays as $\gamma_{ir}$ ($\S$~\ref{standard}) and the
self absorption frequency is solved for using Equation~\ref{nua}.
After $\nu_{cr}$ falls below the observing band, the ejecta emission
turns off, and the observed flux is due to off axis emission that
decays approximately as $t^{-(4+p)/2}$ \citep{kp00}.

\subsection{Inverse Compton cooling by an external source}
We have also allowed for external sources of flux to influence the
cooling of electrons in the ejecta.  In particular, we include
synchrotron flux from the FS incident on the ejecta in our calculation
of $\nu_{cr}$.  The external synchrotron flux in the shocked ejecta
comoving frame is given by
\begin{equation} \label{fex}
f_{ex} = {\tau_{es,ex} c {B'_{ex}}^2 \gamma_{e,ex}^2 \over 6 \pi}
\end{equation}
where the subscript ``\emph{ex}'' denotes values from a source
external to the cooling calculation being done (here, FS flux is the
external flux incident on the RS ejecta), $\gamma_{e,ex}^2$ is the
average electron thermal electron Lorentz factor squared in the
forward shock.  The equation for $\gamma_c$ which includes the
contribution of $f_{ex}$ to the electron cooling is:
\begin{equation} \label{gamcexternal}
\gamma_c = {1 \over \chi_1\left(1+Y\right) + {\chi_2 f_{ex} \over
    \sigma_{KN,ex}}}
\end{equation}
where $\chi_1 = \sigma_T B'^2 t' / \left(6 \pi m_e c^2\right)$,
$\chi_2 = \sigma_T t' / \left(m_e c^2\right)$, and $\sigma_{KN,ex} =
\left[h \nu_{p,ex} \gamma_c / \left(m_e
c^2\right)\right]^{1+\alpha_{ex}}$ is an approximation to the
Klein-Nishina correction on the external flux, and $\nu_{p,ex}$ is the
peak of the $\nu f_{\nu}$ spectrum in the FS; $\alpha_{ex} = -0.5$ if
$\nu_{i,ex} > \nu_{c,ex}$, and $(1-p)/2$ if $\nu_{i,ex} >
\nu_{c,ex}$. We also allow for FS synchrotron self-Compton flux to
influence RS cooling; however the high energy ($\gg m_e c^2$) of the
FS synchrotron flux suppresses self-inverse Compton scattering, thus
not significantly contributing to the RS cooling calculation. We do
not consider the effect of RS emission on FS cooling; at most it is an
order unity effect.

\subsection{Absorption of RS Photons in the FS Material}
We include the effect of absorption of RS photons in the FS
material.  For a selected observing band, $\nu_{obs}$, the
optical depth to absorption is
\begin{equation} \label{tauabsfs}
\tau_{abs,FS} = \left\{ \begin{array}{ll}
  \left({\nu_{obs} \over \nu_{af}}\right)^{5/3} & \nu_{obs} <
  \min(\nu_{cf},\nu_{if})\\
  \left({\nu_{obs} \over \nu_{af}}\right)^{-(p+4)/2} &  \nu_{obs} >
  \min(\nu_{cf},\nu_{if}) \\
  \end{array} \right.
\end{equation}
where $\nu_{af}$ is the observer frame FS synchrotron self absorption
frequency.  Before deceleration, the shocked ejecta and FS medium are
moving together; however, after deceleration, these regions are moving
at different Lorentz factors, and one must take care to use the
appropriate value of $\nu_{af}$ when the RS photons are passing
through the FS.

After calculating the RS emission as described above, the absorption
in the FS is taken into account by reducing the RS flux by a factor of
$\exp(-\tau_{abs,FS})$.  This exponential cut off in flux can
significantly reduce the RS emission, especially for observations at
longer wavelengths, such as the radio.

\subsection{Lepton-enriched ejecta}
We include $e^{\pm}$ pairs in the calculation by simply adding a
certain number of pairs per ejecta proton, $N_{\pm}$, to the
calculation described above.  We do not calculate the creation of
these pairs; we make the assumption that a certain number of pairs per
proton are present in the ejecta already, either being intrinsic from
the source or being created by dissipative processes during the GRB,
prior to the afterglow stage.

We change the radiation calculation described above by multiplying the
column density of electrons by the number of ejecta pairs, $\L \equiv
1+2 N_{\pm}$ (the 1 is for the electrons already accompanying the
protons in the baryonic ejecta) and by dividing the minimum electron
energy by the number of pairs (\ie the new minimum thermal Lorentz
factor is $\gamma_{ir}/\L$).  The dynamics calculation is only altered
when $N_{\pm} \gta m_p/m_e$.  To account for the presence of a high
number of pairs, we alter the dynamics calculation by reducing the
number of protons in the ejecta for a fixed burst energy by $\L_{ep}
\equiv 1 + 2 N_{\pm} m_e/m_p$ (see Table~1).

The RS minimum electron thermal Lorentz factor averaged over the
entire lepton population at shock crossing is
\begin{equation}
\gamma_{ir,+} = { \epsilon_{er}' \left< e_{p,+} \right> \L_{ep} \over
    m_e c^2 \L}.
\end{equation}
Since the RS is only mildly relativistic in the spreading ejecta width
case, the addition of pairs to the ejecta quickly drops $\gamma_{ir}$
into the Newtonian regime.  The already low injection frequency in the
reverse shock can be reduced dramatically by even modest $\L$ (see
Table~1).  The injection frequency of course cannot drop below the
cyclotron frequency; we keep track of this in our numerical
calculation.

The cooling frequency has very weak dependence on $N_{\pm}$ (there is
dependence on $N_{\pm}$ through $\L_{ep}$, but this only makes a
difference if $N_{\pm} \gta 1000$).  If $Y \gg 1$ and $\nu_{ir} <
\nu_{cr}$, $\nu_{cr} \propto \L^{2(p_r-2)/(4-p_r)}$ (for $N_{\pm} \ll
m_p/m_e$).  The dependence of $\nu_{ar}$ on the pair content when
$\nu_{ir} < \nu_{ar} < \nu_{cr}$ is $\nu_{ar} \propto
\L^{(4-2p_r)/(4+p_r)}$ for $N_{\pm} \ll m_p/m_e$; $\nu_{ar}$ is also
fairly insensitive to pair content.

\section{Results} \label{cooling}
In this section, we describe the effects of IC cooling by flux
generated in the RS and FS, absorption in the FS, and the effect of
lepton-enriched ejecta on the RS emission, then determine if there are
observable signatures of pair enriched ejecta.  To ascertain the
effect of each of these new additions to our calculation over the
entire parameter space and a wide range of RS strengths for $s=0,2$,
we randomly vary each parameter in the ranges: $50 \le \Gamma_0 \le
1000$, $1 s \le t_{GRB} \le 100 s$, $0.1 \le E_{52} \le 1000$,
$10^{-3}$ cm$^{-3} \le n_0 \le 100$ cm$^{-3}$ ($10^{-2} \le A_* \le
10$ for $s=2$), $10^{-5} \le \epsilon_{Br} \le 0.1$, $0.01\le
\epsilon_{ir}\le 0.1$, $2.01\le p_r \le 2.91$, and $1 \le N_{\pm} \le
1000$ for 1000 test cases (synthetic GRB afterglows).  All of the
parameters are assumed to have uniform distributions in log space with
the exception of $p_r$, which is assumed to have a uniform
distribution.  The microphysics parameter ranges were chosen
to be consistent with values found from late time afterglow
modeling. 

\subsection{Inverse Compton \& External Cooling} 
\begin{figure}
 \includegraphics[width=0.49\textwidth]{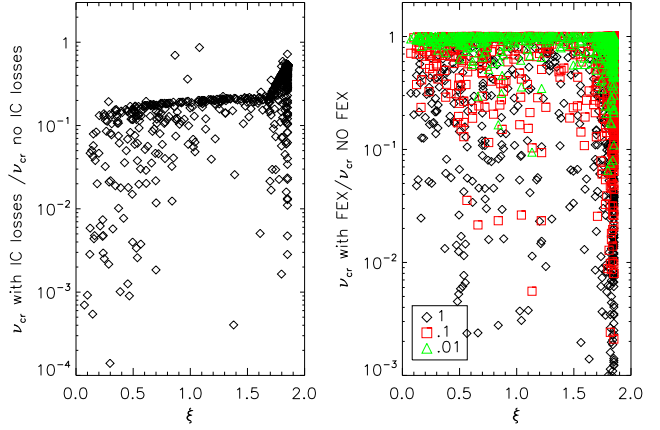}
 \caption{Left panel: Ratio at deceleration of numerically calculated
 $\nu_{cr}$ (with IC cooling) to analytic estimate given in Table 1
 (not including IC cooling) for 1000 test cases in $s=0$. The
 inclusion of IC cooling reduces $\nu_{cr}$ by a factor of roughly 10
 over a wide range of RS strengths ($\xi$).  Right panel: Ratio at
 deceleration of numerically calculated cooling frequency without
 external cooling included to numerical value with external cooling
 included, for the also for the $s=0$ case. The legend gives the value
 of $\epsilon_{Bf}$ in comparison to the value of $\epsilon_{Br}$,
 e.g. for the red squares, $\epsilon_{Bf} = 0.1 \epsilon_{Br}$.}
 \label{fexfigure}
\end{figure}

In the left panel of Figure~\ref{fexfigure}, the ratio of our numerical
value of $\nu_{cr,+}$ to the analytical value given in Table~1 for 1000
test cases is plotted against $\xi$, the dimensionless RS strength
parameter, at deceleration, for $s=0$.  \footnote{$\xi =
(l/\Delta)^{1/2} \Gamma_0^{-4/3}$, where $l = (3 E/(4 \pi n_0 m_p
c^2))^{1/3}$; $\xi \ll 1$ is relativistic RS and $\xi \lta 1$ is the
generic, mildly relativistic RS case.  For $\xi \sim 2$, the ejecta is
in the spreading regime with a mildly to non-relativistic RS (these
points are bunched very closely together in the figures, since $\xi$ is
virtually constant over the dynamics parameter space for all spreading
cases), whereas the non spreading ejecta case ranges from $\xi \ll 1$ to
$\xi \lta 2$.}  Each RS parameter is randomly varied in the ranges
described above, except for $N_{\pm}$, which is set to 0.  For each of
the 1000 cases, the spreading condition is evaluated at deceleration,
and the numerically calculated $\nu_{cr,+}$ is compared to the
analytical value in the proper ejecta evolution case.

The numerically calculated value of $\nu_{cr,+}$, which includes the
self consistent inverse Compton and synchrotron self absorption
calculation as described in $\S$\ref{ICabs}, is in general about a
factor of 10 lower than the analytic value (not including IC losses)
that is typically used over a wide range of RS strengths.  For those
cases where the numerical value is even smaller, the Compton $Y$
parameter is rather large (these cases are more abundant for the non
spreading case, with relativistic RS).  For $s=2$ (not shown in the
figure), the analytical value of $\nu_{cr,+}$ can also be less than
the numerical value of $\nu_{cr,+}$.  This is because our cooling
calculation has taken into account the fraction of the synchrotron
flux that has been synchrotron self-absorbed, reducing the rate of
energy loss for electrons; synchrotron self-absorption contributes
more to the RS cooling calculation in $s=2$ than $s=0$.

External flux influences the RS cooling calculation by decreasing
$\nu_{cr}$.  If the FS radiation was produced with $\epsilon_{Bf}
\sim \epsilon_{Br}$, the external flux can lower $\nu_{cr}$ via IC
scattering by up to a few orders of magnitude for both $s=0,2$ cases.
The effect of external flux IC cooling on the RS synchrotron flux near
deceleration in most cases is relatively small; the effect is greatest
in the x-ray band, where $\nu_{obs} > \left(\nu_{ar},\nu_{ir},
\nu_{cr}\right)$, and flux here can be decreased by a factor of a few
(IC flux is also an important contributing factor to the total light
curve in the x-ray, and is not considered here).  The more important
effect of the external flux is to decrease the cooling frequency below
the R band on a shorter time scale after deceleration, causing the
light curve to fall off very steeply, as $\sim 1/t^3$, soon after
deceleration.  Overall, the effect of external flux on electron
cooling makes RS synchrotron emission a little fainter at the peak of
the light curve and fall off faster after deceleration, making the RS
synchrotron emission more difficult to detect.

In the right panel of Figure~\ref{fexfigure}, we show the effect of
adding external cooling to our calculation.  We plot the ratio of the
numerically calculated $\nu_{cr,+}$ without external cooling to
$\nu_{cr,+}$ including external cooling from the FS. We let $p_f =
p_r$, $\epsilon_{if} = \epsilon_{ir}$, and try three cases of
$\epsilon_{Bf} = (1, 0.1, 0.01) \epsilon_{Br}$.  With external cooling
included, $\nu_{cr}$ in either the spreading or non spreading case can
be reduced by up to 3-4 orders of magnitude!  This is true over a wide
range of RS strengths, and also for $s=2$.  The effect is strongest
when $\epsilon_{Bf} = \epsilon_{Br}$ and weakest when $\epsilon_{Bf} =
0.01 \epsilon_{Br}$.

\subsection{Effect of Absorption in FS} \label{absFS}
Absorption of RS synchrotron photons in the FS is most important in
the radio, as the FS self absorption frequency lies in this band.  In
a large part of the parameter space, the low frequency ($\sim$4 GHz)
RS flux can be completely absorbed; in a another run of 1000 different
RS cases (varying the parameters as described above, and setting FS
$\epsilon_{if}$, $\epsilon_{Bf}$, and $p_f$ equal to RS values), we
find that in approximately 15\% of 1000 test cases, the RS radio (8.5
GHz) flux is completely absorbed by the FS material.  In approximately
40\% of the test cases, $\tau_{abs,FS} >1$ (including those cases
where the RS radio flux is completely absorbed), meaning that the RS
radio flux was decreased by at least a factor of 3.  If we set
$\epsilon_{Bf} = \epsilon_{Br}/100$, we find that 12\% of 1000 test
cases are completely absorbed in the FS, and in 23\% of the cases,
$\tau_{abs,FS} > 1$.  Absorption in the FS may turn out to be a
contributing factor to the difficulty in observing RS flux in the
radio.

\subsection{RS Emission with Pair-Enriched Ejecta} \label{pairSection}
Adding a certain number of pairs per proton to the ejecta adds another
parameter to the RS flux calculation, bringing the total number of
parameters in the RS to four, viz. $\epsilon_{ir}$, $\epsilon_{Br}$,
$p_r$, and $N_{\pm}$\footnote{FS microphysics parameters and the
energy in the explosion, external density, and initial jet opening
angle can be determined, as has been done in the past, by late-time FS
light curve fitting when the RS makes a small contribution to the
total flux.}.  We need four independent measurements of the RS emission
to constrain these four parameters.  Flux in the optical \& X-ray at
early times, spectral index and radio flux at $\sim 0.1$ day are four
quantities that can be observed with the current generation of
instruments.

For $\nu_{obs}>(\nu_{ir},\nu_{cr},\nu_{ar})$ (x-ray) with $Y \ll 1$ or
$\nu_{ir}<\nu_{ar}<\nu_{obs} <\nu_{cr}$ (optical) for any $Y$, the
observed flux at deceleration has a dependence on $N_{\pm}$ of
\begin{equation} \label{fnuLdepxray}
f_{\nu} \propto {\L^{\left(2-p_r\right)} \over
\L_{ep}^{5\left(2-p_r\right)/4}}.
\end{equation}
For $p_r \sim 2$, this flux depends very weakly on the number of pairs
in the ejecta.  For $p_r>2$, the x-ray and optical flux are decreased
from the $N_{\pm} = 0$ case as $N_{\pm}$ is increased (for $N_{\pm}
\ll m_p/m_e$), but the shape of the light curve does not change
(i.e. no breaks are introduced by reducing the injection frequency
below the low value at which it already sits in the absence pairs in
the ejecta).

\begin{figure}
 \includegraphics[width=0.49\textwidth]{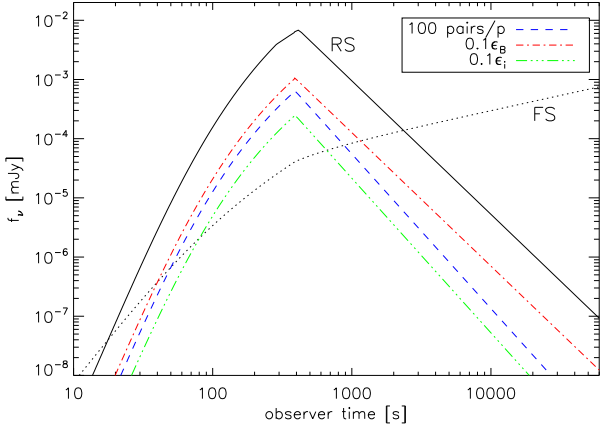}
 \caption{Radio (8.5 GHz) light curve for $s=0$ with the input
 parameters $E_{52}=3$, $\Gamma_0 = 160$, $t_{GRB}=4$ s, $n_0=1.6
 \times 10^{-3}$ cm$^{-3}$, $\epsilon_{Br} = \epsilon_{Bf} = 4 \times
 10^{-5}$, $\epsilon_{ir} =\epsilon_{if} = 0.09$, $p_r = p_f= 2.5$,
 $N_{\pm} = 0$, and $z=1$.  External cooling and absorption in the FS
 are not included in this calculation.  The solid line represents the
 RS light curve with the above parameters and the FS emission is shown
 by the dotted line. The three colored lines alter the parameters in
 the calculation as detailed in the legend.}
 \label{radioPairsfigure}
\end{figure}

In the radio, the observer flux for $\nu_{obs} < \nu_{ir} < \nu_{ar} <
\nu_{cr}$ is proportional to $\L^{-1} \L_{ep}$.  The flux in the radio
is the most greatly affected by the addition of pairs to the
calculation, so the best possible place to look for a signature of
pair-enriched ejecta is in the radio wavelengths. In
Figure~\ref{radioPairsfigure}, we plot an example of RS and FS radio
light curves (8.5 GHz) without pairs (for burst parameters see figure
caption) compared to three cases: (1) adding 100 pairs/p to the
ejecta, (2) decreasing $\epsilon_{Br}$ by a factor of 10, and (3)
decreasing $\epsilon_{ir}$ by a factor of 10.  All three of these
actions produce very similar RS radio light curves; the situation is
highly degenerate, and it would be difficult to distinguish the effect
of pairs in the ejecta on the emission from a variation in the
microphysics parameters in the RS.

1000 RS test cases (parameters varied over ranges described above with
no external cooling or absorption in the FS) show that the peak
frequency of $f_{\nu}$, at
$\max\{\nu_{ar},\min(\nu_{ir},\nu_{cr})\}$, at deceleration also has
very little, if any, dependence on the number of pairs in the ejecta
for $s=0,2$.  This is also true if external cooling is included in the
calculation.  This is in contradiction to the conclusion that
\citet{li03} came to; this is because, over a large portion of the
parameter space, the peak of the spectrum in both the baryonic and
pair-enriched cases is at $\nu_{ar}$, and $\nu_{ar}$ has little
dependence on the number of pairs in the ejecta.

Thus, we conclude there is no single wavelength light curve signature
that depends only on the pair content of the ejecta, and not on the
microphysics parameters in the RS as well.

\subsection{Using Optical and Radio Observations to Determine Pair
 Content}
\label{chipairs}
The best way to discriminate between baryonic and pair-enriched ejecta
is by comparing flux at low frequencies (radio band) and the optical
band.  NP04 suggested a particular combination of the ratio of
observed RS optical and radio fluxes at the peak of each respective
light curve and the ratio of the observer times at which these peaks
occur as a way of determining if the radiation is from the reverse
shock.  We use a similar combination to decide whether the ejecta is
pair enriched.

The parameter we use to determine the pair-enrichment ($\chi$) is
slightly different than that in NP04 (note the difference in exponents
on the ratio of observer times),
\begin{equation} \label{ftratio}
\chi \equiv \left( {F_* \over F_0}\right) \left({t_* \over
  t_0}\right)^k =
    \left({\nu_{opt}\over\nu_{radio}}\right)^{\left(p_r-1\right)\over2}
     \sim 1000
\end{equation}
where
\begin{equation}
k \equiv \left\{ \begin{array}{lc}
{{5\left(p_r-1\right) \over 6} + {9 \over 8}} & s=0 \\
\left(p_r-1\right) +{5\over6} & s=2\\
\end{array} \right.
\end{equation}
and are determined analytically using decay indices after deceleration
of $\nu_{ir} \propto t^{-5/3}$ and $F_{pr} \propto t^{-9/8}$ for $s=0$
and $\nu_{ir} \propto t^{-2}$ and $F_{pr} \propto t^{-5/4}$ for $s=2$
(assuming spreading ejecta evolution for both cases). As in NP04, $t_0$
is the observer time peak of the optical RS emission, $t_*$ is the
observer time peak of the radio RS emission (the peak is produced when
$\nu_{ar}$ falls below the radio), and $F_0$ \& $F_*$ are the observed
fluxes at these two points.  The shell width at deceleration in
both shell evolution cases is $\sim R_+/\Gamma_+^2$ which is $ \ll
R$\footnote{This is true in the non spreading case because of
causality--the distance traveled by the shock front in the shell
comoving frame is of order $R_+/\Gamma_+$, which is $R_+/\Gamma_+^2$ in
the lab frame}, so time delay between photons arising from the front and
back ends of the shell has small effect on the observed flux.

\begin{figure}
 \includegraphics[width=0.49\textwidth]{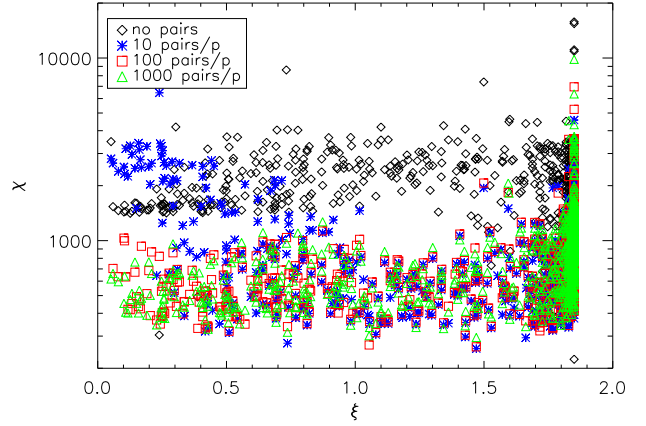}
 \caption{$\chi$ for $p_r = 2.5$ and $s=0$ for 1000 test RS numerical
 light curves, excluding absorption of RS photons in the FS and
 external cooling.  We randomly vary all RS parameters but $p_r$ and
 $N_{\pm}$; other values of $p_r$ give qualitatively similar results
 when adding leptons to baryonic ejecta, however $\chi$ for the
 reference purely baryonic case is higher for higher $p_r$ and vice
 versa. $s=2$ has qualitatively similar results which are discussed in
 $\S$\ref{chipairs}.  The scatter in each case is caused for the most
 part by the variation in dynamics parameters. There are some points
 with very low $\chi$ cut off of the bottom of the plot; these are
 caused when $\nu_{ar} > \nu_{cr}$, and $\nu_{ar}$ tracks $\nu_{cr}$
 in time (which evolves more quickly, leading to smaller $t_*$). }
 \label{fig:np04pairsfigure}
\end{figure}

We plot $\chi$ in Figure~\ref{fig:np04pairsfigure} versus the
dimensionless RS strength parameter, $\xi$, in a run of 1000 RS cases
as described in $\S$~\ref{cooling}, except that $p_r$ is held fixed at
2.5 and $N_{\pm}$ is held constant for each case.  External cooling
and absorption in the FS were turned off. For $N_{\pm} = 0$, the value
for $\chi$ can be larger or smaller by a factor of a few from 1000 for
a constant value of $p_r \sim 2.5$, and is fairly constant over a
large range of reverse shock strengths.  There are a few points which
are cut off the bottom of the plot with abnormally low $\chi$--these
points are caused by $\nu_{cr}$ falling below $\nu_{ar}$.  When this
happens, $\nu_{ar}$ tracks $\nu_{cr}$ (since there are no electrons
radiating above $\nu_{cr}$ after the RS reaches the rear of the
shell), which evolves more quickly than $\nu_{ar}$ does when $\nu_{ar}
< \nu_{cr}$, making $t_*$ and $F_*$ smaller.  This scenario is easily
detected by checking the decay of the radio light curve after the
peak; if the decay is steep, like $\sim 1/t^3$, the abnormally low
value of $\chi$ is due to $\nu_{cr}$ falling below the radio band with
$\nu_{ar}$.

When pairs are added to the ejecta, we find that for $N_{\pm}\gta
100$, the value of $\chi$ has dropped by a factor of about 5 for the
parameter space with $\xi \lta 1.6$ (ultra-relativistic to mildly
relativistic RS).  For $N_{\pm} = 10$, the value for $\chi$ ranges
from $\sim 3000$ when the RS is relativistic ($\xi \ll 1$) to $\sim
200$ when the RS is mildly to non relativistic, making it difficult to
distinguish between baryonic and pair enriched ejecta.  Compared with
the range of $\chi$ for baryonic ejecta (greater or less than 2000 by
a factor of about 2), the drop in $\chi$ by a factor of 5 for $N_{\pm}
\gta 100$ could possibly be used as a tool to distinguish the pair
content of the ejecta.  So, if observations are available for a burst
at the peak of the R band and radio RS emission, and $p_r$ is able to
be determined from spectra, we can calculate $\chi$ and determine if
$N_{\pm} \gta 100$ (it is not possible to determine the exact number
of pairs per proton, but only if $N_{\pm} \gta 100$).  It will be
extremely difficult to tell if the ejecta has a pair content of
$N_{\pm} \lta 100$.

\begin{figure}
 \includegraphics[width=0.49\textwidth]{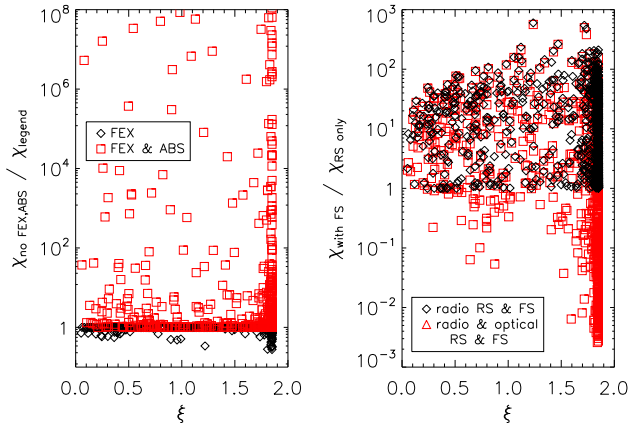}
 \caption{Left panel: Ratio of $\chi$ (no pairs) without external flux
 or absorption in the FS to $\chi$ with external flux and absorption
 in the FS plotted against the value of $\xi$ at deceleration. Note
 that the ratio including absorption can go to arbitrarily high
 numbers, since radio flux can be completely absorbed ($\chi=0$ with
 absorption in FS in these cases); some of these points have been cut
 off the plot.  Right panel: Ratio of $\chi$ calculated including
 contributions of FS in just the radio and in both the radio and
 optical band to $\chi$ calculated without FS contribution, plotted
 against $\xi$ at deceleration.}
 \label{fig:np04exabs}
\end{figure}

In Figure~\ref{fig:np04exabs}, in the left panel, $\chi$ is shown
including the effect of adding external cooling and FS absorption to
the purely baryonic ejecta case.  We repeat the calculation of 1000
test cases with $p_r=2.5$ and $N_{\pm}=0$, and we set $p_f = p_r$,
and vary $\epsilon_{if}$ and $\epsilon_{Bf}$ in the same ranges as the
corresponding RS parameters. The inclusion of external cooling in the
calculation does not affect $\chi$ much.  The value of $\chi$ is
increased at most by a factor of a few from the calculation without
external cooling included, and still lies within the scatter for the
baryonic case shown in Figure~\ref{fig:np04pairsfigure}.

Including absorption in the FS, however, will invalidate
Equation~\ref{ftratio}, since we assumed in deriving it that the RS
radio flux at the light curve peak is equal to the flux at the peak of
the RS spectrum at $t_*$.  This decrease in $F_*$ by a large factor
also decreases $\chi$ by the the same factor ($\chi$ in this case
could be $\lta 1$).  In $\S$\ref{absFS}, we found that absorption in
the FS can be a large effect for much of the parameter space at 8.5
GHz.  Absorption in the FS for observer frequencies less than $\sim
10$ GHz may indeed make it more difficult to accurately determine the
pair content of the ejecta.  However, at higher frequencies, $\sim$
100 GHz, FS absorption is less important and one can calculate $\chi$
at these frequencies to determine $N_{\pm}$ (note that $\chi$ will be
a factor of a few smaller at higher radio frequencies for the purely
baryonic case).

To determine if $\chi$ for $s=2$ has similar properties to the $s=0$
case, we carry out a numerical simulation of 1000 RS test cases for
$s=2$ identical to that done for $s=0$ (without external cooling or
absorption), finding qualitatively similar results as the $s=0$ case.
With $N_{\pm} \lta 100$, the pair content is difficult to determine,
but for $N_{\pm} \gta 100$, the value of $\chi$ is about a factor of 5
lower than in the baryonic case.  In the purely baryonic case, $\chi
\sim 3000 $ for $s=2$ and this value ranges between a factor of 3
higher and lower. The spread of $\chi$ values here is larger
because the range of $n_0 = AR^{-s}$ for $s=2$ is larger than the
range of $n_0$ chosen for the $s=0$ case. 

%the exponent $k$ for the $s=2$ was determined using the decay
%indices of spreading ejecta (as was done in the $s=0$ case), however
%the non spreading ejecta case can become more important in $s=2$.

Up to this point, we have not considered the contaminating effect of
the FS emission to the optical or radio RS emission at the RS light
curve peaks.  It may be most difficult to distinguish RS from FS flux
in observed radio light curves, so we calculate the value of $\chi$
with the FS radio flux included.  In the right panel of
Figure~\ref{fig:np04exabs}, we've plotted the ratio of $\chi$
calculated (for $s=0$) with the FS radio flux contribution to $\chi$
without FS flux at the time which the RS radio light curve peaks.
Again, for 1000 test cases, we set $p_f = p_r$, set $N_{\pm} = 0$, and
vary $\epsilon_{if}$ and $\epsilon_{Bf}$ in the same ranges as the
corresponding RS parameters.  We find that the value of $\chi$ can
increase by a factor of up to 100 from the RS only case, outside of
the scatter in $\chi$. In 60\% of the 1000 test cases, the FS radio
contribution was greater than the RS radio contribution at the RS
radio peak.  For $s=2$, this occurs in 83\% of the 1000 test cases.
One needs to separate the contribution of the RS and FS to the
radio flux using late time data in order to use this tool reliably to
determine the ejecta pair content.

If the RS radio peak was observed, but the light curve has significant
FS contribution, it may be possible to separate the contribution of FS
from the RS radio flux by continued monitoring of the radio band for a
period of a week or so, when the FS radio LC peaks.  This information
can be used to determine the contribution of the FS to the radio flux
at the time of the RS peak, and $\chi$ due to the RS alone. If the
radio flux is dominated by the FS contribution during the RS peak and
the RS peak is not observed, then this tool cannot be used to
determine ejecta pair content.

We also find that the optical RS peak may be as difficult to observe
as the radio peak due to FS contamination, in agreement with
the conclusion drawn in NP04.  In the right panel of
Figure~\ref{fig:np04exabs}, we also show the ratio of $\chi$ with FS
contribution (both radio and optical) to $\chi$ with only RS flux.
The inclusion of FS optical flux at deceleration increases the scatter
in $\chi$ even more; $\chi$ with FS flux ranges from being 100 times
smaller to 100 times larger larger than $\chi$ with RS flux only.  For
the $s=0$ case, the FS optical flux at deceleration is brighter than
the RS optical flux 70\% of 1000 test cases; in the $s=2$ case, this
occurred in only 30\% of 1000 test cases.  This is another difficulty
in calculating $\chi$ for observed light curves.  However, as in the
radio, if an RS optical peak is observed, the FS contribution may be
able to be removed if the optical light curve is followed for long
enough after deceleration to determine the FS contribution to the
total observed flux. We note that the fact that the FS optical
emission is larger than the RS optical emission at deceleration in
$\sim$70\% of test cases for $s=0$ may explain the lack of optical
flashes and rapidly declining light curves at detected early times.

Scintillation in the radio may also be a problem for observing the RS
radio peak due to the high level of variability that this process
introduces into the observed radio light curve, as was observed in GRB
970508 \citep{frail97,taylor97}.  The fluctuations are more pronounced
at early times, when the RS radio light curve is expected to peak,
because the source size is smaller.  One can reduce fluctuations from
scintillation by observing at frequencies higher than 8.5 GHz.
Observations made at frequencies near 50 GHz and even in the
millimeter range (250 GHz) may be more suitable for determining $\chi$
and $N_{\pm}$ because of the insensitivity to scintillation at these
frequencies.  The effect of absorption in the FS is also smaller for
these higher energy radio photons ($\tau_{abs,FS} > 1$ for a much
smaller parameter space), and the RS light curve peaks at earlier
times at these frequencies when the FS light curve may not be as
bright.  With current technology, radio observations on the
timescales necessary for calculating $\chi$ of few minutes/hours to
days after the burst are feasible at around 8.5 GHz.  At higher
frequencies like the sub-millimeter, recent observations are typically
being done $\sim$ 0.1 to 1 day post-burst \citep{smith05}; the
timescale for the sub-millimeter light curve peak is from a few
minutes to a few hours after the burst (up to about 0.1 day).

Other things that may limit the usefulness of $\chi$ as a tool for
estimating $N_{\pm}$ include sparseness of sampling of the radio light
curve near the RS and FS peaks and breaks in the electron energy
distribution between the optical and radio.  The sparseness of radio
data points near the RS and FS light curve peaks may introduce an
error into the determination of $\chi$ of a factor of a few.
\citet{pk02} have found evidence in late-time afterglow modeling of a
break in the electron energy distribution between the radio and the
optical; one must be careful to choose the correct value of $p_r$
(value of $p_r$ between $\nu_{ir}$ and the optical band) for the
calculation of $\chi$.

In summary, it is possible to use optical and radio flux and observer
times at the respective peaks of the light curves to determine pair
content if the number of pairs/proton is $\gta 100$.  There are many
factors, however, that one must take into account when using this
tool.  The contributions to the optical and radio flux from the RS and
FS must be separated, which may be difficult and requires good time
coverage from $\sim$ 1 minute to hours in the optical and $\sim$ 1
hour to days in the radio.  In the radio band, below 10 GHz,
absorption in the FS and interstellar scintillation pose problems, and
in order to avoid these issues, one should use observations at a
higher frequency than $\sim 100$ GHz.

\subsection{Constraining Parameters with Available Observations and
  Upper Limits}

\begin{figure}
\includegraphics[width=0.49\textwidth]{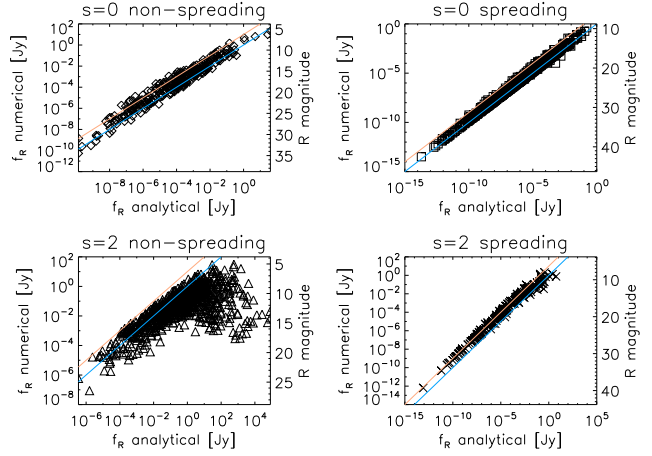}
\caption{Comparison of 1000 test cases of numerically calculated
  R-band flux at deceleration with analytical expression for each case
  of $s=0,2$ spreading/non-spreading. All parameters, including
  $N_{\pm}$, are varied. The full analytical expressions assume $
  (\nu_{ir},\nu_{ar}) < \nu_{obs} < \nu_{cr}$ and are given in the
  last row of Table~1 for the case of $p=2.5$, $z=1$ and in Appendix A
  for the general case.  A linear relationship (blue line) and $f_R$
  analytical $=$10 times $f_R$ numerical (orange line) are shown on
  each panel for a guide.  }
\label{fRfullfig}
\end{figure}

\begin{figure}
 \includegraphics[width=0.49\textwidth]{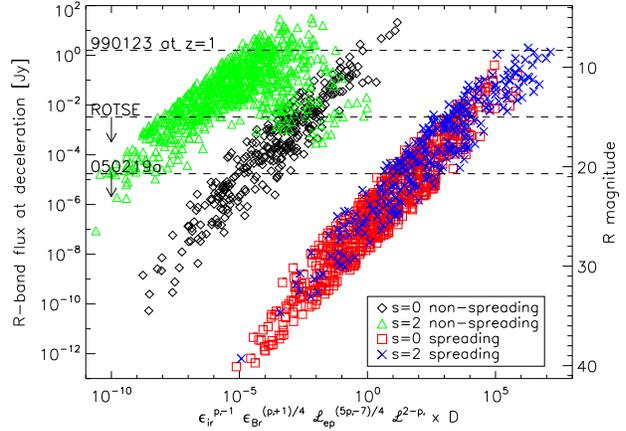}
 \caption{Observer frame RS flux in the $R$-band at deceleration
 assuming $z=1$ and including pairs (no external cooling or absorption
 in the FS) plotted against the parameter dependences in the
 expression for flux in the R band at deceleration (without
 constants).  All parameters are varied. The variable $D$ in the
 expression on the x-axis contains all of the dynamics parameters and
 is different for each of the four cases shown: $D= {n_0^{p_r/4}
 \Gamma_0^{p_r-2} E_{52}^{5/4} t_{dur}^{-3/4}}$ for $s=0$
 non-spreading, $ {A_*^{p_r/2} \Gamma_0^{p_r-2} E_{52}^{(5-p_r)/4}
 t_{dur}^{-(3+p_r)/4}}$ for $s=2$ non-spreading, ${n_0^{(p_r+1)/4}
 \Gamma_0^{p_r} E_{52}}$ for $s=0$ spreading, and $ {A_*^{3(1+p_r)/4}
 \Gamma_0^{1+2 p_r} E_{52}^{(1-p_r)/2}}$ for $s=2$ spreading.  The
 three dashed lines denote the observed $R$ band flux at deceleration
 for GRB 990123 shifted to $z=1$, observed upper limits reported in
 \citet{kehoe01} from the ROTSE telescope, and the $V$-band upper
 limit reported by {\emph Swift} UVOT for GRB 050219a
 \citep{schady05}, shown for comparison to expected theoretical $R$
 band flux from the RS. Note that 990123 was an exceptional burst--a
 very small fraction of the parameter space produces optical flashes
 this bright.}
 \label{fig:fnuRfigure}
\end{figure}

In Figure~\ref{fRfullfig}, we plot the comparison of our numerically
calculated R band flux at deceleration with the full analytical
expression given in the last row of Table~1 ($p=2.5$ and $z=1$) and in
Appendix A (full expression).  The numerical calculation does not
include external cooling or absorption in the FS (the latter is
unimportant for the R band, but the former could decrease the flux by
a factor of $\sim 2$), and all parameters are varied, including
$N_{\pm}$.  The analytical expressions assume $ (\nu_{ir},\nu_{ar}) <
\nu_{obs} < \nu_{cr}$ and include pairs. For each case of $s=0$ or
$s=2$, we ran 1000 test cases, and separated these 1000 cases into two
sets determined by their ejecta width at deceleration.  The blue line
shown on each plot shows a linear relationship for a guide, and the
orange line shows the relationship if the numerical flux were 10 times
the analytically calculated flux.

In all four cases, the scatter around the linear relationship is due
to the estimations made for $\Gamma_{ej}$ and more importantly
$\gamma_{ir}$ in the analytical expression.  $\gamma_{ir}$ is more
difficult to estimate in the non spreading case, since the range of RS
strengths ranges from ultra-relativistic to Newtonian.  Since we use a
mildly relativistic RS estimate for the analytical expression, the
analytical expression underestimates the ultra-relativistic RS flux and
overestimates the Newtonian RS flux.  Likewise, in the spreading case,
we made an estimate of $\gamma_{ir}$ being mildly relativistic; all
spreading cases are mildly relativistic to Newtonian, so the scatter
around a linear relationship is much smaller than for non spreading
ejecta.  There is still a small bit of scatter due to the value of
$\gamma_{ir}$ that we chose (see Table~1), and the analytical value is
typically within a factor of a few to 10.

The analytical expression given in Table~1 is a good estimation of the
R band flux at deceleration for both spreading cases.  It is less
useful for the non spreading cases; in the $s=2$ non-spreading case,
the ordering of break frequencies assumed in the analytical expression
is often not applicable, and the linear relationship does not hold
over the entire parameter space explored--this occurs in the top right
hand corner of the plot for this case.  The cooling frequency is
typically lower in $s=2$ compared to $s=0$, and drops below the other
two break frequencies frequently in this calculation.

In Figure \ref{fig:fnuRfigure}, we have done the same calculation as
described above for Figure~\ref{fRfullfig}, only here we have plotted
the numerical observer frame R band flux against only the parameter
dependences from the analytical expression for R band flux.  For
example, for $s=0$, $\Delta=R_+/\left(2\Gamma_0^2\right)$, we have
plotted numerically calculated RS flux $f_{\nu}$ against the
analytical combination of parameters $\epsilon_{ir}^{p_r-1}
\epsilon_{Br}^{(p_r+1)/4} \L^{2-p_r} n_0^{(p_r+1)/4} \Gamma_0^{p_r}
E_{52}$.  The scatter (width) of the spreading regions on this plot
are mainly due to the variation of $p_r$, while the scatter of the non
spreading regions is due to $p_r$ and the dynamics parameters.  The
$s=2$ non spreading case is again not quite linear, for reasons
discussed above for Figure~\ref{fRfullfig}.

This plot can be used as a tool to constrain the burst parameters
using observed R-band flux or upper limits at or near deceleration
using the more accurate results of the numerical calculation. Also,
this plot can be used to view the ranges of expected R-band flux at
deceleration if the early afterglow is caused by the RS with baryonic
or pair enriched ejecta.  Lines are drawn onto the plot to compare to
the early afterglow detection of GRB 990123 at the peak (scaled from
$z=1.61$ to $z=1$) and upper limits from ROTSE and \emph{Swift}.

We find that optical flash emission from GRB 990123 falls at the very
bright end of the distribution of RS R band flux at deceleration for
all 4 cases shown in Figure~\ref{fig:fnuRfigure}.  There are a great
many synthetic bursts in our calculation, especially in the case of
$s=0$, spreading ejecta, with intrinsically low R band RS flux levels
for burst parameters consistent with those found from afterglow
modeling.  This may indeed be the reason that the growing number of
GRBs with rapid follow up do not show a bright optical flash.

As an example of how to use Figure~\ref{fig:fnuRfigure} to constrain
burst parameters with optical data or upper limits near deceleration,
we look at the case of GRB 050219a.  GRB050219a was a 23.6 s long {\it
Swift}-detected burst of with a fluence of $\left(5.2 \pm 0.4\right)
\times 10^{-6}$ erg cm$^{-2}$ and was detected in the x-ray by XRT
\citep{romano05}.  An optical upper limit from {\it Swift} of $V =
20.7$ was found 96 s after the burst \citep{schady05}.  We find that
the upper limit for this burst falls near the bottom of the
distribution of the non-spreading cases (although the distribution can
be extended to lower values of $f_{\nu,R}$ using even lower limits for
the ranges on $n_0$/$A_*$ and $\epsilon_{Br}$) but is in the middle of
the distribution for the two spreading cases.  It seems that RS flux
fainter than this upper limit is fairly typical in the case of
spreading ejecta.

Using Figure~\ref{fig:fnuRfigure}, we can make constraints on the
burst parameters of 050219a (assuming that the upper limit of $V=20.7$
applies at deceleration for this burst and there was no
extinction). In the $s=2$ cases, we can make fairly severe constraints
on $\epsilon_{Br}$ and $A_*$.  For example, using the plot for the
$s=2$ non spreading case, if we assume $z=1$ (although see
\citealt{berger05} for the redshift distribution for {\it Swift} GRBs
so far) and $p_r \sim 2.2$ we find (see Table~1)
\begin{equation} \label{plotLimit}
\epsilon_{ir}^{1.2} \epsilon_{Br}^{0.8}
      \L_{ep} \L^{-0.2} A_*^{1.1} \Gamma_0^{0.2} E_{52}^{1.8}
  t_{dur}^{-1.3} \hspace{2 mm} \lta \hspace{2 mm} 1.2 \times 10^{-8}.
\end{equation}
Using the observed burst duration and using the assumption $z=1$,
$t_{dur} = 11.8$ s, and estimating $E_{52}$ from the fluence, we set
$E_{52} \sim 2$.  We set $\epsilon_{ir} = 0.1$, a typical value, as
indicated from afterglow modeling \citep{pk02}.  Inserting these
values and setting $\Gamma_0 = 100 \Gamma_{0,2}$, we have
\begin{equation}
\epsilon_{Br}^{0.8}
\L^{-0.2} A_*^{1.1} \Gamma_{0,2}^{0.2} \hspace{2 mm} \lta
\hspace{2 mm} 5.3 \times 10^{-7}.
\end{equation}
Since the dependence of this relation on the number of pairs and
$\Gamma_{0,2}$ is small, we can say that within a factor of a few,
$A_*^{1.1} \epsilon_{Br}^{0.8} \lta 5.3 \times 10^{-7} $, implying
very small values for $A_*$ and/or $\epsilon_{Br}$.

If we set $\epsilon_{Br} \sim 10^{-5}$, on the lower end of the
distribution of values found from afterglow modeling \citep{pk02},
$A_* \lta 10^{-2}$.  Also, if we choose to make $\epsilon_{ir}$
smaller for this burst than the typical value we have chosen, say
$\epsilon_{ir} \sim 0.01$, then if $A_* \sim 1$, $\epsilon_{Br} \lta
10^{-7}$, still a rather low value.

If we try the $s=2$ spreading case, we find similarly severe
requirements for $A_*$ and $\epsilon_{Br}$.  The expression for the
spreading case, however, is much more sensitive to $\Gamma_{0,2}$, so
the limits on $A_*$ and $\epsilon_{Br}$ are not as robust as in the
non spreading case above.  For $A_* \sim 1$ and $\Gamma_{0,2} \sim 1$,
$\epsilon_{Br} \lta 10^{-9}$.  If we require $\epsilon_{Br} \sim
10^{-5}$, then $A_* \sim 0.1$.  In summary, for the $s=2$ density
profile to apply, with either spreading or non spreading ejecta, the
upper limit at 96 s requires low values of $A_*$ or $\epsilon_{Br}$
near deceleration.  For the $s=0$ cases, using the typical values
above ($\epsilon_{ir} \sim 0.1$), we find the optical upper limit at
96 s to be less constraining on $n_0$ and $\epsilon_{Br}$.

\section{Discussion}
In this paper, we have calculated RS emission from purely baryonic and
pair-enriched ejecta.  We take into account the mildly relativistic
nature of the RS (we allow for the full range of RS speeds, from
Newtonian to ultra-relativistic), and self consistently calculate
synchrotron self absorption and inverse Compton cooling in the ejecta.
Additionally, we have allowed for the ejecta to be cooled by FS
synchrotron flux incident on the ejecta and for the low-energy RS
photons to be absorbed in the FS material as they move outward from
the source.

We find that the flux in the R band at deceleration depends very
weakly on the ejecta pair content.  The RS radio emission is affected
by ejecta pairs the most, however the shape of the radio light curve
is not affected.  The effect of pairs on any single wavelength light
curve (a reduction in the flux with increasing number of pairs per
proton) can be replicated by varying the shock microphysics parameters
in the RS; it is a degenerate problem.  It is impossible to determine
the pair content of the ejecta from a single wavelength light curve.

It may be possible to determine if $N_{\pm} \gta 100$ by using
observations of the RS optical and radio light curve peaks. By
calculating $\chi$ (see Equation~\ref{ftratio}) using observations,
one may be able to determine if $N_{\pm} \gta 100$; one cannot
determine the precise number of pairs per proton with $\chi$, but
whether there are a significant number of pairs present in the ejecta.
The value of $\chi$ for purely baryonic ejecta is fairly constant (for
a given value of $p_r$) over a wide range of RS strengths; effects
that may increase the spread in $\chi$ (reducing the effectiveness of
$\chi$ as a tool for determining ejecta pair content) include
absorption of radio RS photons in the FS, contribution of the FS
emission to the RS flux at the light curve peaks, and scintillation in
the radio light curve.  External (FS synchrotron) flux incident on the
ejecta has little effect on the value of $\chi$.  By using radio
observations at higher frequencies, e.g. at 50 GHz or 250 GHZ (mm),
effects of scintillation and FS contribution to radio emission can be
reduced, and $\chi$, and therefore the pair content of the ejecta, can
be more accurately determined.

\citet{li03} have also looked at the emission from pair enriched
ejecta; we agree that it is possible for a large number of pairs to be
present in the ejecta and that the resulting R band flux is largely
insensitive to the number of pairs in the ejecta, however we come to
different conclusions regarding the resulting ejecta emission.
\citet{li03} predict a strong flash in the IR band; we find that the
peak frequency of the RS is largely unaffected by the pair content of
the ejecta.  Indeed, the injection frequency is reduced greatly by
even a modest number of pairs, however we find that in the most
frequent arrangement of frequencies, the peak of $f_{\nu}$ is at
$\nu_{ar}$, which is not affected by pairs much at all.  Most of the
difference between our calculations result from the treatment of RS
strength; we have taken into account the mildly relativistic nature of
the reverse shock, where \citet{li03} have approximated the RS as
being highly relativistic. Another contributing factor to the
difference in our results is the more careful and self consistent
calculation of synchrotron cooling and self absorption frequencies
that we have done here.

In conclusion, RS emission from pair-enriched ejecta looks very
similar to that from purely baryonic ejecta.  The most promising tool
to determine if GRB ejecta is pair-enriched, by the measurement of the
parameter $\chi$ defined in Equation~\ref{ftratio}, depends greatly on
well sampled observations near the peak of the light curves in the
optical and the radio; if the optical flash is seen quickly after the
burst with {\it Swift} or other ground based telescopes, radio follow
up in frequency bands between 50 GHz and 250 GHz (mm) from 15 minutes
to $\sim$ 1 day should provide some information about the pair
enrichment of the ejecta.

This research was supported in part by grants from NSF (AST-0406878)
and NASA-Swift (NNG05G185) (PK) and the US-Israel BSF (TP).

\appendix
\section{R band flux at deceleration}
Here we write the full analytic expressions for the R band flux at
deceleration (in cgs units) for each of the four cases in Figure~5.  
\begin{displaymath}
f_{\nu,R} = {\sigma_T \epsilon_{ir}^{p_r-1} \epsilon_{Br}^{(p_r+1)\over4}
      \L_{ep}^{(5p_r-7)\over4} \L^{2-p_r}
      \left(1+z\right)^{(1-p_r)\over2} \over
      H_0^2 \left( \sqrt{1+z}-1\right)^2 }\times
\end{displaymath}
\begin{equation}
\left\{ \begin{array}{ll}
{0.018 \left(5.2 \times
      10^{16}\right)^{(1-p_r)\over2}
      m_p^{(5p_r-8)\over4} n_0^{p_r\over4} \Gamma_0^{p_r-2}
      E_{52}^{5\over4}  \over  m_e^{(3p_r-5)\over2} q^{(3-p_r)\over2}
      \pi^{(6+p_r)\over4} c^{9\over4} t_{dur}^{3\over4}} &  s=0
\hspace{4 pt} c
t_{dur} \\
0.086 \left(5.8\times
  10^{14}\right)^{(1-p_r)\over2} A_*^{p_r\over2} \Gamma_0^{p_r-2}
  E_{52}^{(5-p_r)\over4}
  m_p^{(3p_r-4)\over2}  
  \over m_e^{(3p_r-5)\over2}
  q^{(3-p_r)\over2} \pi^{3\over2} c^{(9-p_r)\over4}
  t_{dur}^{(3+p_r)\over4}  &  s=2 \hspace{4 pt} c
t_{dur} \\
{0.03 \left(1.1 \times 10^{18}\right)^{(1-p_r)\over2}
  n_0^{(p_r+1)\over 4} \Gamma_0^{p_r} E_{52} m_p^{(5p_r-7)\over4}
  \over m_e^{(3p_r-5)\over2} \pi^{(5+p_r)\over4} q^{(3-p_r)\over2} c
  }& s=0 \hspace{4 pt} {R \over 2 \Gamma_0^2} \\
{4.06 \left(7.7 \times 10^{16}\right)^{(1-p_r)\over2} c^{p_r}
  A_*^{3(1+p_r)\over 4}  \Gamma_0^{1+2 p_r} 
  m_p^{(7p_r-5)\over4} 
  \over m_e^{(3p_r-5)\over2} \pi^{(3-p_r)\over4} q^{(3-p_r)\over2}
  E_{52}^{(p_r-1)\over 2} } & s=2
\hspace{4 pt} {R \over 2 \Gamma_0^2} 
\end{array} \right.
\end{equation}

\end{document}